# Mobility Improvement and Temperature Dependence in MoSe$_2$ Field-Effect Transistors on Parylene-C Substrate

Bhim Chamlagain[1],Qing Li[2], Nirmal Jeevi Ghimire[3,4], Hsun-Jen Chuang[1], Meeghage Madusanka Perera[1], Honggen Tu[5], Yong Xu[5], Minghu Pan[2], Di Xaio[6], Jiaqiang Yan[3,4], David Mandrus[3,4] and Zhixian Zhou[1, a),]

[1]Department of Physics and Astronomy, Wayne State University, Detroit, MI 48201
[2] Center for Nanophase Materials Sciences, Oak Ridge National Laboratory, Oak Ridge, TN 37831
[3]Deptment of Materials Science and Engineering, The University of Tennessee, Knoxville, TN 37996
[4]Materials Science and Technology Division, Oak Ridge National Laboratory, Oak Ridge, TN 37831
[5]Department of Electrical and Computer Engineering, Wayne State University, Detroit, MI 48202
[6]Department of Physics, Carnegie Mellon University, Pittsburg, PA

a) Author to whom correspondence should be addressed, electronic mail: zxzhou@wayne.edu

## Abstract

We report low temperature scanning tunneling microscopy characterization of MoSe$_2$ crystals, and the fabrication and electrical characterization of MoSe$_2$ field-effect transistors on both SiO$_2$ and parylene-C substrates. We find that the multilayer MoSe$_2$ devices on parylene-C show a room temperature mobility close to the mobility of bulk MoSe$_2$ ( 100 cm$^2$V$^{-1}$s$^{-1}$ − 160 cm$^2$V$^{-1}$s$^{-1}$ ), which is significantly higher than that on SiO$_2$ substrate ($\approx$50 cm$^2$V$^{-1}$s$^{-1}$). The room temperature mobility on both types of substrates are nearly thickness independent. Our variable temperature transport measurements reveal a metal-insulator transition at a characteristic conductivity of e$^2$/h. The mobility of MoSe$_2$ devices extracted from the metallic region on both SiO$_2$ and parylene-C increases up to $\approx$ 500 cm$^2$V$^{-1}$s$^{-1}$ as the temperature decreases to $\approx$ 100 K, with the mobility of $MoSe_2$ on $SiO_2$ increasing more rapidly. In spite of the notable variation of charged impurities as indicated by the strongly sample dependent low temperature mobility, the mobility of all $MoSe_2$ devices on $SiO_2$ converges above 200 K, indicating that the high temperature ( > 200 K) mobility in these devices is nearly independent of the charged impurities. Our atomic force microscopy study of $SiO_2$ and parylene-C substrates further rule out the surface roughness scattering as a major cause of the substrate dependent mobility. We attribute the observed substrate dependence of $MoSe_2$ mobility primarily to the surface polar optical phonon scattering originating from the $SiO_2$ substrate, which is nearly absent in $MoSe_2$ devices on parylene-C substrate.

The successful isolation of two-dimensional (2D) graphene has stimulated research on a broad range of other 2D materials, among which layered transition metal dichalcogenides (TMDs) have attracted particular attention.[1-15] Similar to graphene, atomic layers of covalently bonded chalcogen-metal-chalcogen units can be extracted from bulk TMD crystals by a mechanical cleavage technique due to the relatively weak van der Waals interactions between the layers. The semiconducting members of the TMD family including $MoS_2$, $MoSe_2$, $WS_2$ and $WSe_2$ have not only demonstrated many of the "graphene like" properties highly desirable for electronic applications such as a relatively high mobility, mechanical flexibility, chemical and thermal stability, and the absence of dangling bonds, but also have a substantial band gap (1 ~ 2 eV depending on the material and its thickness), which is absent in 2D graphene but required for mainstream logic applications.[1, 16-18] For example, in contrast to the low ON/OFF ratios in graphene field-effect transistors (FETs), an ON/OFF ratio of $> 10^8$ has been reported in monolayer $MoS_2$.[5]

Despite these recent progress, the mobility values of monolayer and multilayer $MoS_2$ devices on $SiO_2$ (the most commonly used substrate for $MoS_2$ FETs) reported by multiple groups were substantially below the Hall mobility of bulk $MoS_2$ (100~200 $cm^2V^{-1}s^{-1}$),[3, 7, 9-10] which greatly hinders their application potential in multifunctional electronic devices.  In addition to the intrinsic scattering from phonons in the TMD channel,[15] the carrier mobility of TMD transistors on $SiO_2$ is expected to be also limited by extrinsic scattering from charged impurities at the channel/substrate interface and charge traps in $SiO_2$, substrate surface roughness, and remote surface optical phonons originating from $SiO_2$. Coulomb scattering from charged impurities at the channel/substrate interface has been proposed as the predominant cause of the relatively low room-temperature mobility in $MoS_2$.[7, 19] Although a high-$\kappa$ dielectric may screen Coulomb scattering from charged impurities, complete recovery of the intrinsic phonon-limited mobility has not been observed in high-$\kappa$ dielectric encapsulated $MoS_2$ devices. On the other hand, the presence of a low energy optical phonon mode in $SiO_2$ (~ 60 meV) may also cause non-negligible surface polar optical scattering[20] and significantly reduced the mobility, as suggested by a recent theoretical study.[21] However, experimental investigations of surface polar optical phonon effects on the channel mobility of TMD FETs are still lacking. In order to tap into the full potential of TMDs as a channel material for high performance FETs, it is crucial to use substrate/dielectric materials that do not further reduce the TMD mobility *via* surface polar optical phonon scattering.

In this article, we present a detailed temperature dependent electrical study of ultrahigh crystalline quality multilayer $MoSe_2$ FETs of varying thickness (5-15 nm) on $SiO_2$ and parylene-C substrates. Parylene-C, a cross-linkable polymer widely used as a passivation layer and gate dielectric, is an excellent candidate for substrate because its lowest energy optical phonon mode of 130 meV (corresponding to the vibrational stretch of C-Cl bond) cannot be easily excited at room temperature.[22-23] $WSe_2$ FETs with parylene top gate dielectric have demonstrated high room temperature mobility up to 500 $cm^2$/Vs.[24] Moreover, parylene-C is insoluble in common solvents such as acetone and isopropanol and thus comparable with the standard device fabrication process. Our four-terminal electrical measurements of multiple $MoSe_2$ FETs on $SiO_2$ reveal that the room temperature mobility is nearly thickness independent with an average value of $\approx 50$ $cm^2V^{-1}s^{-1}$ in good agreement with previously reported results from $MoSe_2$ and $MoS_2$ on $SiO_2$ substrates.[25-26] However, a two to three-fold mobility improvement is consistently observed in $MoSe_2$ FETs on parylene-C substrate compared to $MoSe_2$ devices on $SiO_2$. To elucidate the origin of the strong substrate dependence of mobility, we measured the electrical characteristics of $MoSe_2$ on both types of substrates between 77 K and 295 K. As the temperature decreases, the temperature dependence of the mobility for $MoSe_2$ FETs on both types of substrates behave as $\mu \sim T^{-\gamma}$, indicating that the charge transport is dominated by phonon scattering in both cases. However, the mobility of $MoSe_2$ FETs on $SiO_2$ substrate increases more rapidly (larger $\gamma$) than the devices on parylene-C substrate with decreasing temperature, before eventually merging at $\mu \approx 500 \ cm^2V^{-1}s^{-1}$ and $T \approx 100$ K. The stronger temperature dependence of $MoSe_2$ devices on $SiO_2$ than on parylene-C can be primarily attributed to an additional surface polar optical phonon scattering contribution originating from the $SiO_2$ substrate, which is consistent with the substantially lower room temperature mobility in $MoSe_2$ devices on $SiO_2$ than on parylene-C. Our study reveals the important role of surface polar phonon scattering in carrier mobility and demonstrates parylene-C as an excellent substrate for TMD FETs.

**RESULTS AND DISCUSSIONS**

$MoSe_2$ crystals were synthesized by chemical vapor transport using iodine as transport agent. The as grown crystals were phase pure as determined by x-ray diffraction. To further characterize the quality of the $MoSe_2$ crystals, scanning tunneling microscopy (STM) measurements were performed on freshly cleaved surfaces of $MoS_2$ crystals inside an ultra-high vacuum (UHV) chamber at 4.5 K without any additional thermal treatment to avoid any possible thermally induced surface reconstruction. Figure 1a shows a representative STM topographic image of cleaved $MoSe_2$ surface measured by 10 nm × 10 nm, where the atomically resolved honeycomb structures bare close resemblance to other layered systems such as $MoS_2$ and graphene. The 1×1 unit cell is shown as a rhombus in the high resolution image (Figure 1b) with a lattice distance of 3.3 Å expected for $MoSe_2$.[27] Remarkably, the surface within a relatively large scan area of 10 nm by 10 nm is surprisingly clean and nearly defect/impurity free (see Figure 1a). Since defects and/or impurities reduce the mean free path of the charge carriers and thus the mobility by serving as scattering centers for charge transport, the extremely low impurity level and high crystalline quality of our $MoSe_2$ is critical to achieving its ultimate materials and device performance.[14, 28]

For electrical transport studies, thin $MoS_2$ crystals (5 nm - 15 nm thick) were produced by repeated splitting of bulk crystals using a mechanical cleavage method, and subsequently transferred to degenerately doped silicon substrates covered either by 290 nm $SiO_2$ or by 130 nm parylene-C vapor deposited on top of 290 nm $SiO_2$.[3-4, 29] Optical microscopy was used to identify thin $MoSe_2$ crystals, which were further characterized by non-contact mode atomic force microscopy (AFM). We chose 5 nm - 15 nm thick $MoSe_2$, because multilayer $MoSe_2$ of this thickness range has a much high yield of sufficiently large flakes (for patterning multiple electrodes) than thinner samples and a relatively smaller $c$-axis interlay resistance compared to thicker samples.[30] $MoSe_2$ FET devices were fabricated using standard electron beam lithography and electron beam deposition of 5 nm of Ti and 50 nm of Au.[31] To eliminate electrical contact contributions, we also patterned voltage probes in between drain and source electrodes to facilitate four-terminal measurements. A schematic illustration and an AFM image of typical $MoSe_2$ devices are shown in Figure 1c and 1d, respectively. Electrical properties of the devices were measured by a Keithley 4200 semiconductor parameter analyzer in a Lakeshore Cryogenic probe station under high vacuum (~ $1×10^{-6}$ Torr).

Figure 2a and 2b depict the room temperature transfer characteristics of two ~12 nm MoSe$_2$ devices on SiO$_2$ and parylene-C, respectively. Both devices exhibit highly asymmetric ambipolar behavior, with the ON/OFF current ratio exceeding $10^6$ for electrons and less than $10^3$ for holes at a drain-source voltage of 1 V. The asymmetry between electron and hole transport may be attributed to  1) a relatively large Schottky barrier height for the hole channel as the Fermi level of the contact metal ( Ti ) tends to line up much closer to the conduction band edge than the valence band edge in MoSe$_2$, and  2) small amount of intrinsic *n*-doping in the transport channel.[32]  The hysteresis in the transfer characteristics is likely due to the charge injection from the adsorbates (such as moisture and oxygen) on the channel surfaces and/or at the interfaces between the channel and the substrate.[33] The hysteresis could be reduced by sweeping the gate voltage in a smaller range as discussed below. In future studies, the observed hysteresis could also be removed using a glove box as achieved by multiple groups.[34-36] As shown in Figure 2c and 2d, the drain-source current of both devices is linear at low drain-source voltages.  As $V_{ds}$ increases, the drain-source current starts to saturate in the low gate voltage range ( $V_{bg} < 10$ V and  $< 30$ V for devices on SiO$_2$ and parylene-C, respectively), while remaining linear at higher gate voltages. The current saturation at low gate voltages is likely caused by the reduction of the effective $V_{bg}$ and $V_{ds}$ due to the relatively large parasitic series drain/source contact resistance ($R_C$) given by $V_{bg\_eff} = V_{bg} - R_C I_{ds}$ and $V_{ds\_eff} = V_{ds} - 2R_C I_{ds}$.[15] At higher $V_{bg}$, the contact resistance is lowered by the reduction of the effective Schottky barrier height through band bending, leading to more linear $I_{ds}$-$V_{ds}$ behavior signifying near Ohmic contacts.[37]

To investigate the true channel-limited electronic performance of MoSe$_2$ devices and understand the charge transport mechanisms, particularly the role of the substrate, we measured the back gate dependence of conductivity σ for MoSe$_2$ devices both on SiO$_2$ and on parylene-C in a four-terminal configuration. The four-terminal conductivity is defined as $\sigma = I_{ds} \times \frac{L}{W}/V_{inn}$, where $L$ and $W$ are the separation between the voltage probes (the inner contacts) and sample width, respectively; and $V_{inn}$ is the measured voltage difference between the voltage probes (kept below 50 mV in all measurements). Field-effect mobility is extracted from the $V_{bg}$ dependence of $\sigma$ using the expression $\mu = 1/C_{bg} \times d\sigma/dV_{bg}$ in the linear region of the $\sigma$ *vs* $V_{bg}$ curves, where $C_{bg}$ is the back-gate capacitance per unit area. Based on a simple parallel plate capacitor model, $C_{bg}$ is determined to be $1.2 \times 10^{-8}$ F cm$^{-2}$ for 290 nm SiO$_2$ ($C_{bg} = 3.9 \times \varepsilon_0/290$ nm) and $7.6 \times 10^{-9}$ F cm$^{-2}$ for 130 nm parylene-C on 290 nm SiO$_2$ ($C_{bg} = \frac{3.9 \times 3.12 \times \varepsilon 0}{130\ nm \times 3.9 + 290\ nm \times 3.12}$), respectively.[29] Figure 3a shows the room temperature conductivity as a function of back gate voltage for two representative MoSe$_2$ devices: a 10 nm thick MoSe$_2$ on SiO$_2$ and a 12 nm thick MoSe$_2$ on a parylene-C . In spite of the qualitatively similar transfer and output characteristics between devices on SiO$_2$ and parylene-C ( Figure 2 ), the mobility of the MoSe$_2$ device on parylene-C ($\approx$ 118 cm$^2$V$^{-1}$s$^{-1}$) is significantly larger than that on SiO$_2$ ($\approx$ 50 cm$^2$V$^{-1}$s$^{-1}$).

To eliminate any possible sample-to-sample variations, we systematically measured 11 MoSe$_2$ devices on SiO$_2$ and 5 MoSe$_2$ devices on parylene-C in the four-terminal configuration. Figure 3b plots room temperature field-effect mobility as a function of channel thickness for all measured MoSe$_2$ FETs devices, with thickness ranging from $\sim$ 5 nm to 14 nm. The mobility of our SiO$_2$-supported MoSe$_2$ devices slightly fluctuates around an average value of $\approx$50 cm$^2$V$^{-1}$s$^{-1}$ without showing any noticeable thickness dependence, consistent with previous results from SiO$_2$-supported multilayer MoS$_2$.[21] In spite of the extremely high crystalline quality of our MoSe$_2$ samples (as shown in Figure 1a), the average room temperature mobility of our MoSe$_2$ devices on SiO$_2$ substrate is rather low compared to the Hall mobility of bulk MoSe$_2$ (about 100

cm$^2$V$^{-1}$s$^{-1}$ − 200 cm$^2$V$^{-1}$s$^{-1}$), but similar to that observed in SiO$_2$-supported MoS$_2$ devices fabricated from commercially available MoS$_2$ crystals.[28,26] This suggests that the room temperature mobility of our MoSe$_2$ devices on SiO$_2$ is likely limited by extrinsic scattering mechanisms. In contrast, the room temperature mobility of all five MoSe$_2$ devices on parylene-C ranges from ≈100 cm$^2$V$^{-1}$V$^{-1}$s$^{-1}$ to ≈150 cm$^2$V$^{-1}$V$^{-1}$s$^{-1}$, which is close to the bulk values.[28] Since the error bars in the mobility data are mainly caused by the uncertainties in the channel length between the voltage probes due to their finite width, the greater absolute fluctuations in the mobility on parylene-C is directly related to the higher mobility on parylene-C.

To understand the substrate/dielectric dependent mobiltiy in our MoSe$_2$ devices, we consider various mobility-limiting scattering mechanisms as formulated by Mathiessen's rule $\mu^{-1} = \mu_{INT}^{-1} + \mu_{SD}^{-1} + \mu_{CI}^{-1} + \mu_{SR}^{-1} + \mu_{SPP}^{-1}$. Here $\mu_{INT}^{-1}$ represents the mobility limited by intrinsic scattering from lattice phonons, $\mu_{SD}^{-1}$ presents mobility limited by structural defects, and $\mu_{CI}^{-1}$, $\mu_{SR}^{-1}$ and $\mu_{SPP}^{-1}$ represent mobility limited by extrinsic scattering from Coulomb impurities (CI), the surface roughness (SR), and the dielectric surface polar optical phonons (SPP), respectively. The intrinsic mobility $\mu_{INT}$ and mobility limited by structural defects $\mu_{SD}$ are expected to be similar for MoSe$_2$ FETs both on SiO$_2$ and on parylene-C, since all our devices are fabricated from the same MoSe$_2$ crystal. We also exclude structural defects as a major source of scattering given the extremely high crystalline quality of our MoSe$_2$ crystals. Next, we consider the effects of surface roughness scattering on the mobility of our MoSe$_2$ devices. Figure 3c and 3d show AFM topographic images acquired in the vicinity of MoSe$_2$ samples on SiO$_2$ and parylene-C, respectively, from which the RMS surface roughness is determined to be 0.3 nm for SiO$_2$ and 0.6 nm for parylene-C. The observation of higher mobility on rougher parylene-C substrate rules out surface roughness scattering as a major cause of the substrate dependent mobility in our MoSe$_2$ devices. Therefore, we can safely deduce that the lower mobility in MoSe$_2$ devices on SiO$_2$ is likely due to additional Coulomb and/or interfacial surface polar optical phonon scattering.

Charged impurities present at the interface between the semiconducting channel and substrate have been generally considered as the primary cause of low room-temperature mobility in TMD devices.[7, 38] Radisavljevic *et al.* recently showed significantly improved mobility in monolayer MoS$_2$ devices with a high-κ HfO$_2$ top-gate dielectric, which was attributed to effective damping of Coulomb scattering on charged impurities.[7] However, the mobility observed in their MoS$_2$ devices (up to 60 cm$^2$V$^{-1}$s$^{-1}$ at 260K) is still much lower than the theoretical values or experimental results from bulk samples.[19, 28] A possible cause of this discrepancy is surface polar optical phonon scattering from substrate and gate dielectric. Indeed, the temperature-dependent mobility curve in reference [7] was recently reproduced by a Monte Carlo method taking into account both the charged impurity scattering and surface polar optical phonon scattering.[21] Considerable mobility improvement was also reported in multiplayer MoS$_2$ on PMMA dielectric in comparison with MoS$_2$ on SiO$_2$, which was attributed to the reduced short-range disorder and long range disorder at the channel/PMMA interface than at the channel/SiO$_2$ interface.[21] However, the lack of temperature dependent mobility data makes it difficult to further ellucidate the origin of substrate dependent mobility in these devices.

To shed additional light on the origin of the significant mobility difference between MoSe$_2$ devices on SiO$_2$ and on parylene-C, we systematically measured the temperature dependence of four-terminal conductivity *vs* gate voltage. Figures 4a and 4b show the temperature dependence of four-terminal conductivity as a function of back gate voltage for two representative MoSe$_2$ devices: one on SiO$_2$ and one on parylene-C. At low gate voltages, both devices show typical insulating or semiconducting behavior with the conductivity increasing with temperature. At high gate voltages, the conductivity decreases with increasing temperature, characteristic of metallic behavior. To further examine the crossover from an insulating or semiconducting state to a metallic state, we plot the conductivity as a function of temperature at various gate voltages in Figure 4c and 4d. It is apparent that a metal-insulator transition occurs at a critical gate voltage between 30 and 35 V, corresponding to a critical carrier density of $\sim$ $1 \times 10^{12}$ cm$^{-2}$ and $\sim 7 \times 10^{11}$ cm$^{-2}$ for the SiO$_2$ and parylene-C substrates, respectively. Above this critical carrier density, the conductivity decreases with increasing temperature, corresponding to a metallic behavior. Below this critical carrier density, the conductivity increases with increasing temperature, characteristic of insulating or semiconducting behavior. More interestingly, this metal-insulator transition is associated with a critical conductivity of e$^2$/h, consistent with metal-insulator transitions (MITs) observed in monolayer, bilayer and multiplayer MoS$_2$ as well as theoretical expectations for 2D semiconductors.[7, 8, 11] To rule out the possible hysteresis effects on the observed MIT, we also measured the four-terminal conductivity as a function of gate voltage in both the "up" and "down" gate-sweep directions. As shown in Figure 5b, hysteresis is nearly absent in the four-terminal conductivity, while the two terminal conductivity of the same device shows substantial hysteresis. The significantly reduced hysteresis in our four-terminal conductivity data is likely due to the smaller gate sweep range of our four-terminal measurement than two-terminal measurement. Since all our four-terminal conductivity results were consistently measured in a relatively small gate voltage range (-10 V < $V_{bg}$ < 60 V) where the hysteresis is negligibly small, these results were unlikely influenced by possible hysteresis effects.

Figure 5a shows the temperature dependence of field-effect mobility for six $MoSe_2$ devices on $SiO_2$ (solid symbols) and three $MoSe_2$ devices on parylene-C (hollow symbols) extracted from the linear region of the conductivity curves in the metallic state ($35 < V_{bg} < 45V$), using the expression for field-effect mobility $\mu = 1/C_{bg} \times d\sigma/dV_{bg}$. The mobility values of all three devices on parylene-C follow a $\mu \sim T^{-\gamma}$ dependence with $\gamma \approx 1.2$ for the entire measured temperature range. This is consistent with the theoretical modeling of phonon-limited mobility in layered TMD materials such as $MoS_2$ and $MoSe_2$, which shows $\mu \sim T^{-\gamma}$ dependence where the exponent $\gamma$ depends on the dominant phonon scattering mechanism with $\gamma \sim 2.4$ in bulk $MoSe_2$ samples and lower for carriers in 2D.[19, 28] The relatively low $\gamma$ in our devices suggests that the charge carriers are likely confined in 2D and behave as a 2D electron gas, which is consist with the recent finding of Li *et al.* that the carriers in a 14 layer $MoS_2$ FET (about 10 nm thick) are largely confined within 1-2 nm range near the interface of the gated dielectric.[30] A similar $\mu \sim T^{-\gamma}$ dependence with a higher $\gamma \approx 1.7$ is observed in two of the six $MoSe_2$ devices on $SiO_2$ (5.2 nm and 14 nm thick, respectively), indicating that their mobility is also predominantly limited by phonon scattering. The mobility of other four $MoSe_2$ device on $SiO_2$ also follows the same phonon limited behavior (with the same power exponent $\gamma \approx 1.7$) at temperatures above 160 K. As the temperature decreases from 160 K to 77 K, their mobility starts to saturate, likely limited by Coulomb scattering due to the varying amount of charged impurities at the $MoSe_2/SiO_2$ interface as previously observed in $MoS_2$ devices.[7, 10] Higher $\gamma$ value for $MoSe_2$ devices on $SiO_2$ than on parylene-C suggests additional temperature dependent scattering mechanism(s) in $SiO_2$-supported $MoSe_2$ devices.

A likely scenario is that the mobility in $MoSe_2$ on $SiO_2$ is further reduced by additional polar optical phonon scattering from the underlying $SiO_2$. The $SiO_2$ surface polar optical phonon mode with an energy of ≈ 60 meV can be easily excited by thermal energy at room temperature, while such a soft surface polar optical phonon mode is absent in parylene-C. [20, 22] At about 100 K, the phonon limited mobility values for $MoSe_2$ devices on both $SiO_2$ (those with lower level of charged impurities) and parylene-C merge, which is expected as the mobility undergoes a transition from being dominated by optical phonon scattering (including surface polar optical phonons) to being dominated by acoustic phonon scattering at ∼ 100 K.[19] Another possible source of stronger temperature dependence of the mobility observed in $MoSe_2$ devices on $SiO_2$ is the presence of greater amount of charged impurities at the $SiO_2/MoSe_2$ interface than at the parylene/MoSe2 interface. Recently, Ong and Fischetti showed theoretically that the increase of mobility with decreasing temperature (behaving as $\mu \sim T^\gamma$), which is commonly interpreted to be a signature of phonon-limited electron transport, could also be limited by CI scattering due to the weakening of charge screening within the TMD channel as the temperature increases.[39] Although CI scattering could contribute to the temperature dependence of the mobility as a separate term, we believe that the lower room temperature mobility and larger γ in our $MoSe_2$ devices on $SiO_2$ than on parylene-C substrate is unlikely to be chiefly caused by stronger CI scattering for the following reasons. Firstly, the mobility of all six $MoSe_2$ devices on $SiO_2$ converges above 200 K in spite of the notable variation of charged impurities as indicated by the strongly sample dependent low temperature mobility (Figure 5), indicating that the high temperature ( > 200 K) mobility in these devices is nearly independent of the charged impurities. Secondly, the mobility values of our devices on both $SiO_2$ and parylene-C are significantly higher than the CI limited mobility from the calculations of Ong and Fischetti, suggesting that CI scattering plays a less significant role in our devices compared to the theory.[39] At carrier densities comparable to the lower end of the carrier density range for the calculation of CI/phonon-limited mobility of bear $MoS_2$ devices in reference [39] ($\approx 10^{12}$ cm$^{-2}$), the mobility of our devices is at least an order of magnitude higher than the calculated mobility. As the carrier density increases, the field-effect mobility in our devices slightly decreases (as indicated by the slight decrease of the slope in the conductivity-gate voltage curves shown in Figure 4a and 4b), while the CI/phonon-limited mobility in reference [39] increases. Thirdly, the temperature dependence of mobility in our devices is qualitatively different from the previously reported CI scattering limited field-effect mobility of $MoS_2$ devices, in which case the mobility decreases with decreasing temperature below 200 K.[7]

In the insulating (or semiconducting) region ($V_{bg} < 35$V), the temperature dependence of conductivity shows a thermally activated behavior as depicted in Figures 6a and 6b for devices on $SiO_2$ and parylene-C, respectively. Since the conductivity was measured in a four-terminal configuration, we exclude Schottky barriers as a possible explanation. An activation energy $E_a$ can be extracted using the express $\sigma \sim \exp\left(-\frac{E_a}{k_B T}\right)$, where $k_B$ is the Boltzmann constant. As shown in Figure 6c, the activation energy $E_a$ decreases with the gate voltage, which can be attributed to the decease of energy gap between the Fermi level $E_F$ and conduction band edge $E_C$ as the Fermi level is tuned toward the conduction band by the gate voltage. The slope of the curves at low gate voltages (when the devices are in the fully depleted region) can be expressed as $\frac{dE_a}{dV_{bg}} = -\frac{dE_F}{dV_{bg}} = -\frac{eC_{bg}}{(C_{bg} + e^2 D(E))}$, where $D(E)$ represents the density of trap states at the interface between the $MoSe_2$ channel and substrate. The density of trap states is found to be $\approx 7.6 \times 10^{12}$ cm$^{-2}$ and $\approx 4.9 \times 10^{12}$ cm$^{-2}$ for $MoSe_2$ on $SiO_2$ and parylene-C, respectively, in excellent agreement with the $D(E)$ of multiplayer $MoS_2$ on $SiO_2$ reported by Ayari *et. al* ($7.2 \times 10^{12}$ cm$^{-2}$).[40] The rather similar $D(E)$ for $SiO_2$ and parylene-C substrates further indicates that CI scattering is unlikely the limiting factor of the drastically different mobility in our MoSe$_2$ devices on SiO$_2$ and parylene-C.

High-κ HfO$_2$ has been favored as a dielectric material for TMD transistors due to its capability to screen charged impurities and its effectiveness in tuning the charge carriers.[5, 7] However, the presence of a soft polar phonon vibration mode in HfO$_2$ along with its high dielectric constant may lead to severe surface polar optical phonon scattering. [21] A thin layer of parylene may be used as a buffer layer between the TMD channel and HfO$_2$ to reduce the surface polar phonon scattering from HfO$_2$ while taking advantage of its high dielectric constant to effectively screen the charged impurities and tune the charge density in the TMD channel.

**CONCLUSIONS**

In conclusion, we demonstrated that the room temperature mobility in multiplayer MoSe$_2$ FETs fabricated on parylene-C approaches its bulk value and is significantly higher than that in MoSe$_2$ devices on SiO$_2$. We attribute the observed mobility difference primarily to the additional surface polar optical phonon scattering originating from the SiO$_2$ substrate but nearly absent in parylene-C. The additional polar optical phonon scattering from SiO$_2$ substrate at the MoSe$_2$/SiO$_2$ interface also leads to a stronger temperature dependence of the mobility in MoSe$_2$ on SiO$_2$ than on parylene-C. At sufficiently low temperatures where acoustic phonons dominate, the mobility of MoSe$_2$ devices both on SiO$_2$ and parylene-C merge. Our variable temperature study of the substrate dependence of the mobility in MoSe$_2$ further demonstrates that substrate surface polar phonons may be a significant limiting factor of room temperature mobility in TMD FETs.

## Methods

Parylene-C was deposited on degenerately doped silicon substrate with 290 nm thermal oxide at room using di-para-xylyene(DPX) as the precursor in a commercially available parylene coating system (PDS 2010). Multilayer $MoSe_2$ flakes were produced by mechanical exfoliation of high quality $MoSe_2$ crystals and subsequently transferred to $Si/SiO_2$ substrates with and without 130 nm of parylene-C. Optical microscopy and Park-Systems XE-70 noncontact mode atomic microscopy (AFM) were used to identify and characterize thin $MoSe_2$ flakes. $MoSe_2$ FET devices were fabricated using standard electron beam lithography and subsequent electron beam deposition of 5nm of Ti covered by 50 nm of Au. Electrical properties of the devices were measured by a Keithley 4200 semiconductor parameter analyzer in a lakeshore Cryogenic probe station under high vacuum ($1\times10^{-6}$ Torr).

*Conflict of interest:* The authors declare no competing financial interest.

*Acknowledgement*

This work was supported by NSF (ECCS-1128297 and DMR-1308436 ). Part of this research was conducted (MP, QL) at the Center for Nanophase Materials Sciences under project # CNMS2011-066, which is sponsored at Oak Ridge National Laboratory by the Scientific User Facilities Division, Office of Basic Energy Sciences, U.S. Department of Energy. NJG, DX, JY and DM were supported by Materials and Engineering Division, Office of Basic Energy Sciences, U.S. Department of Energy.

**Figure Captions**

**Figure 1.** (a) 10 nm×10 nm atomic resolution STM topography ($V_{bias}$= -0.5 V, $I_{set}$= 100 pA) of a cleaved MoSe$_2$ crystal measured at 4.5 K. (b) Close-up image showing a defect-free hexagonal lattice. (c) Schematic illustration of the cross-sectional view of back-gated MoSe$_2$ devices on SiO$_2$ (top) and parylene-C (bottom) with Au/Ti (50 nm/5 nm) contacts and electrical connections for electrical characterization including drain/source electrodes and voltage probes for four-terminal measurements. (d) AFM topography of a typical MoSe$_2$ device.

**Figure 2**. Room temperature transfer characteristics of multilayer MoSe$_2$ FET devices fabricated on (a) SiO$_2$ and (b) Parylene-C substrates. (c,d) Room temperature output characteristics of the devices in (a) and (b), respectively. The MoSe$_2$ samples in both (a) and (b) are ~ 12 nm thick.

**Figure 3**. (a) Room temperature four-terminal conductivity as a function of gate voltage for multilayer MoSe$_2$ FETs fabricated on SiO$_2$ and parylene-C substrates. The MoSe$_2$ samples on SiO$_2$ and parylene-C are 14 nm and 12 nm thick, respectively. (b) Field-effect mobility *versus* MoSe$_2$ thickness extracted from multiple MoSe$_2$ devices fabricated on SiO$_2$ (solid circle) and parylene-C (solid square), where the error bars are mainly caused by the uncertainties in the channel length between the voltage probes due to the finite width of the voltage electrodes. (c,d) AFM images of SiO$_2$ and parylene-C surfaces, respectively.

**Figure 4.** Temperature dependent four-terminal conductivity as a function of gate voltage for MoSe$_2$ devices on (a) SiO$_2$ and (b) parylene-C. (c, d) Gate voltage dependent four-terminal conductivity as a function of temperature for the same devices in (a) and (b), respectively.

**Figure 5**. (a) Temperature dependence of field-effect mobility extracted from the four-terminal conductivity *versus* gate voltage measurements on MoSe$_2$ devices on SiO$_2$ (solid symbols) and parylene-C (hollow symbols) with various thicknesses between 5 nm and 14 nm. (b) Four-terminal and two-terminal conductivity of a $\approx$ 10 nm thick MoSe$_2$ device on parylene-C measured by sweeping the gate from negative to positive and then from positive to negative voltages.

**Figure 6.** Arrhenius plot of conductivity of MoSe$_2$ devices on (a) SiO$_2$ and (b) parylene-C in the insulating region. (c) Dependence of activation energy on gate voltage.

1.      Wang, Q. H.; Kalantar-Zadeh, K.; Kis, A.; Coleman, J. N.; Strano, M. S. Electronics and Optoelectronics of Two-Dimensional Transition Metal Dichalcogenides. *Nat Nano* **2012,** *7*, 699-712.

2.      Butler, S. Z.; Hollen, S. M.; Cao, L.; Cui, Y.; Gupta, J. A.; Gutiérrez, H. R.; Heinz, T. F.; Hong, S. S.; Huang, J.; Ismach, A. F.; Johnston-Halperin, E.; Kuno, M.; Plashnitsa, V. V.; Robinson, R. D.; Ruoff, R. S.; Salahuddin, S.; Shan, J.; Shi, L.; Spencer, M. G.; Terrones, M.; Windl, W.; Goldberger, J. E. Progress, Challenges, and Opportunities in Two-Dimensional Materials Beyond Graphene. *ACS Nano* **2013,** *7*, 2898-2926.

3.      Novoselov, K. S.; Jiang, D.; Schedin, F.; Booth, T. J.; Khotkevich, V. V.; Morozov, S. V.; Geim, A. K. Two-Dimensional Atomic Crystals. *Proc. Natl. Acad. Sci.* **2005,** *102*, 10451-10453.

4.      Novoselov, K. S.; Geim, A. K.; Morozov, S. V.; Jiang, D.; Zhang, Y.; Dubonos, S. V.; Grigorieva, I. V.; Firsov, A. A. Electric Field Effect in Atomically Thin Carbon Films. *Science* **2004,** *306*, 666-669.

5.      Radisavljevic, B.; Radenovic, A.; Brivio, J.; Giacometti, V.; Kis, A. Single-Layer $MoS_2$ Transistors. *Nature Nanotech.* **2011,** *6*, 147-150.

6.      Yin, Z.; Li, H.; Li, H.; Jiang, L.; Shi, Y.; Sun, Y.; Lu, G.; Zhang, Q.; Chen, X.; Zhang, H. Single-Layer $MoS_2$ Phototransistors. *ACS Nano* **2011,** *6*, 74-80.

7.      Radisavljevic, B.; Kis, A. Mobility Engineering and a Metal–Insulator Transition in Monolayer $MoS_2$. *Nat Mater* **2013,** *12*, 815-820.

8.      Baugher, B. W. H.; Churchill, H. O. H.; Yang, Y.; Jarillo-Herrero, P. Intrinsic Electronic Transport Properties of High-Quality Monolayer and Bilayer $MoS_2$. *Nano Lett.* **2013,** *13*, 4212-4216.

9.      Zhang, Y.; Ye, J.; Matsuhashi, Y.; Iwasa, Y. Ambipolar $MoS_2$ Thin Flake Transistors. *Nano Lett.* **2012,** *12*, 1136–1140.

10.     Jariwala, D.; Sangwan, V. K.; Late, D. J.; Johns, J. E.; Dravid, V. P.; Marks, T. J.; Lauhon, L. J.; Hersam, M. C. Band-Like Transport in High Mobility Unencapsulated Single-Layer $MoS_2$ Transistors. *Appl. Phys. Lett.* **2013,** *102*, 173107.

11.     Ye, J. T.; Zhang, Y. J.; Akashi, R.; Bahramy, M. S.; Arita, R.; Iwasa, Y. Superconducting Dome in a Gate-Tuned Band Insulator. *Science* **2012,** *338*, 1193-1196.

12.     Tongay, S.; Zhou, J.; Ataca, C.; Lo, K.; Matthews, T. S.; Li, J.; Grossman, J. C.; Wu, J. Thermally Driven Crossover from Indirect toward Direct Bandgap in 2d Semiconductors: $MoSe_2$ *Versus* $MoS_2$. *Nano Lett.* **2012,** *12*, 5576.

13.     Fang, H.; Chuang, S.; Chang, T. C.; Takei, K.; Takahashi, T.; Javey, A. High-Performance Single Layered $WSe_2$ p-Fets with Chemically Doped Contacts. *Nano Lett.* **2012,** *12*, 3788–3792.

14.     Jones, A. M.; Yu, H.; Ghimire, N. J.; Wu, S.; Aivazian, G.; Ross, J. S.; Zhao, B.; Yan, J.; Mandrus, D. G.; Xiao, D.; Yao, W.; Xu, X. Optical Generation of Excitonic Valley Coherence in Monolayer $WSe_2$. *Nat Nano* **2013,** *8*, 634-638.

15.     Liu, W.; Kang, J.; Sarkar, D.; Khatami, Y.; Jena, D.; Banerjee, K. Role of Metal Contacts in Designing High-Performance Monolayer N-Type $WSe_2$ Field Effect Transistors. *Nano Lett.* **2013,** *13*, 1983-1990.

16.     Bolotin, K. I.; Ghahari, F.; Shulman, M. D.; Stormer, H. L.; Kim, P. Observation of the Fractional Quantum Hall Effect in Graphene. *Nature* **2009,** *462*, 196.

17.     Novoselov, K. S.; Geim, A. K.; Morozov, S. V.; Jiang, D.; Katsnelson, M. I.; Grigorieva, I. V.; Dubonos, S. V.; Firsov, A. A. Two-Dimensional Gas of Massless Dirac Fermions in Graphene. *Nature* **2005,** *438*, 197-200.

18.     Yoon, Y.; Ganapathi, K.; Salahuddin, S. How Good Can Monolayer $MoS_2$ Transistors Be? *Nano Lett.* **2011,** *11*, 3768-3773.

19.     Kaasbjerg, K.; Thygesen, K. S.; Jacobsen, K. W. Phonon-Limited Mobility in N-Type Single-Layer $MoS_2$ from First Principles. *Phys. Rev. B* **2012,** *85*, 115317.

20.     Perebeinos, V.; Rotkin, S. V.; Petrov, A. G.; Avouris, P. The Effects of Substrate Phonon Mode Scattering on Transport in Carbon Nanotubes. *Nano Lett* **2008**, *9*, 312-316.

21.     Zeng, L.; Xin, Z.; Chen, S.; Du, G.; Kang, J.; Liu, X. Remote Phonon and Impurity Screening Effect of Substrate and Gate Dielectric on Electron Dynamics in Single Layer $MoS_2$. *Appl. Phys. Lett.* **2013**, *103*, 113505.

22.     Jakabovič, J.; Kováč, J.; Weis, M.; Haško, D.; Srnánek, R.; Valent, P.; Resel, R. Preparation and Properties of Thin Parylene Layers as the Gate Dielectrics for Organic Field Effect Transistors. *Microelectronics J* **2009**, *40*, 595-597.

23.     Li, Y.; Su, L.; Shou, C.; Yu, C.; Deng, J.; Fang, Y. Surface-Enhanced Molecular Spectroscopy (Sems) Based on Perfect-Absorber Metamaterials in the Mid-Infrared. *Sci. Rep.* **2013**, *3*.

24.     Podzorov, V.; Gershenson, M. E.; Kloc, C.; Zeis, R.; Bucher, E. High-Mobility Field-Effect Transistors Based on Transition Metal Dichalcogenides. *Appl. Phys. Lett.* **2004**, *84*, 3301-3303.

25.     Larentis, S.; Fallahazad, B.; Tutuc, E. Field-Effect Transistors and Intrinsic Mobility in Ultra-Thin $MoSe_2$ Layers. *Appl. Phys. Lett.* **2012**, *101*, 223104.

26.     Bao, W.; Cai, X.; Kim, D.; Sridhara, K.; Fuhrer, M. S. High Mobility Ambipolar MoS[Sub 2] Field-Effect Transistors: Substrate and Dielectric Effects. *Appl. Phys. Lett.* **2013**, *102*, 042104.

27.     Böker, T.; Severin, R.; Müller, A.; Janowitz, C.; Manzke, R.; Voß, D.; Krüger, P.; Mazur, A.; Pollmann, J. Band Structure of $MoS_2$, $MoSe_2$, and A-$MoTe_2$: Angle-Resolved Photoelectron Spectroscopy and Ab Initio Calculations. *Phys. Rev.  B* **2001**, *64*, 235305.

28.     Fivaz, R.; Mooser, E. Mobility of Charge Carriers in Semiconducting Layer Structures. *Phys. Rev.* **1967**, *163*, 743-755.

29.     Sabri, S. S.; Lévesque, P. L.; Aguirre, C. M.; Guillemette, J.; Martel, R.; Szkopek, T. Graphene Field Effect Transistors with Parylene Gate Dielectric. *Appl. Phys. Lett.* **2009**, *95*, 242104.

30.     Li, S.-L.; Wakabayashi, K.; Xu, Y.; Nakaharai, S.; Komatsu, K.; Li, W.-W.; Lin, Y.-F.; Aparecido-Ferreira, A.; Tsukagoshi, K. Thickness-Dependent Interfacial Coulomb Scattering in Atomically Thin Field-Effect Transistors. *Nano Lett* **2013**, *13*, 3546-3552.

31.     Lin, M.-W.; Ling, C.; Agapito, L. A.; Kioussis, N.; Zhang, Y.; Cheng, M. M.-C.; Wang, W. L.; Kaxiras, E.; Zhou, Z. Approaching the Intrinsic Band Gap in Suspended High-Mobility Graphene Nanoribbons. *Phys. Rev.  B* **2011**, *84*, 125411.

32.     Perera, M. M.; Lin, M.-W.; Chuang, H.-J.; Chamlagain, B. P.; Wang, C.; Tan, X.; Cheng, M. M.-C.; Tománek, D.; Zhou, Z. Improved Carrier Mobility in Few-Layer $MoS_2$ Field-Effect Transistors with Ionic-Liquid Gating. *ACS Nano* **2013**, *7*, 4449-4458.

33.     Late, D. J.; Liu, B.; Matte, R. H. S. S.; Dravid, V. P.; Rao, C. N. R. Hysteresis in Single-Layer $MoS_2$ Field Effect Transistors. *ACS Nano* **2012**, *6*, 5635–5641.

34.     Lui, C. H.; Liu, L.; Mak, K. F.; Flynn, G. W.; Heinz, T. F. Ultraflat Graphene. *Nature* **2009**, *462*, 339-341.

35.     Xu, K.; Cao, P.; Heath, J. R. Graphene Visualizes the First Water Adlayers on Mica at Ambient Conditions. *Science* **2010**, *329*, 1188-1191.

36.     Sandilands, L. J.; Shen, J. X.; Chugunov, G. M.; Zhao, S. Y. F.; Ono, S.; Ando, Y.; Burch, K. S. Stability of Exfoliated $Bi_2Sr_2Dy_xCa_{1−X}Cu_2O_{8+\Delta}$ Studied by Raman Microscopy. *Phys. Rev.  B* **2010**, *82*, 064503.

37.     Das, S.; Chen, H.-Y.; Penumatcha, A. V.; Appenzeller, J. High Performance Multilayer $MoS_2$ Transistors with Scandium Contacts. *Nano Lett.* **2012**, *13*, 100-105.

38.     Kim, S.; Konar, A.; Hwang, W.-S.; Lee, J. H.; Lee, J.; Yang, J.; Jung, C.; Kim, H.; Yoo, J.-B.; Choi, J.-Y.; Jin, Y. W.; Lee, S. Y.; Jena, D.; Choi, W.; Kim, K. High-Mobility and Low-Power Thin-Film Transistors Based on Multilayer $MoS_2$ Crystals. *Nature Commun.* **2012**, *3*, 1011.

39.     Ong, Z.-Y.; Fischetti, M. V. Mobility Enhancement and Temperature Dependence in Top-Gated Single-Layer $MoS_2$. *Phys. Rev.  B* **2013,** *88*, 165316.

40.     Ayari, A.; Cobas, E.; Ogundadegbe, O.; Fuhrer, M. S. Realization and Electrical Characterization of Ultrathin Crystals of Layered Transition-Metal Dichalcogenides. *J. Appl. Phys.* **2007,** *101*, 014507.

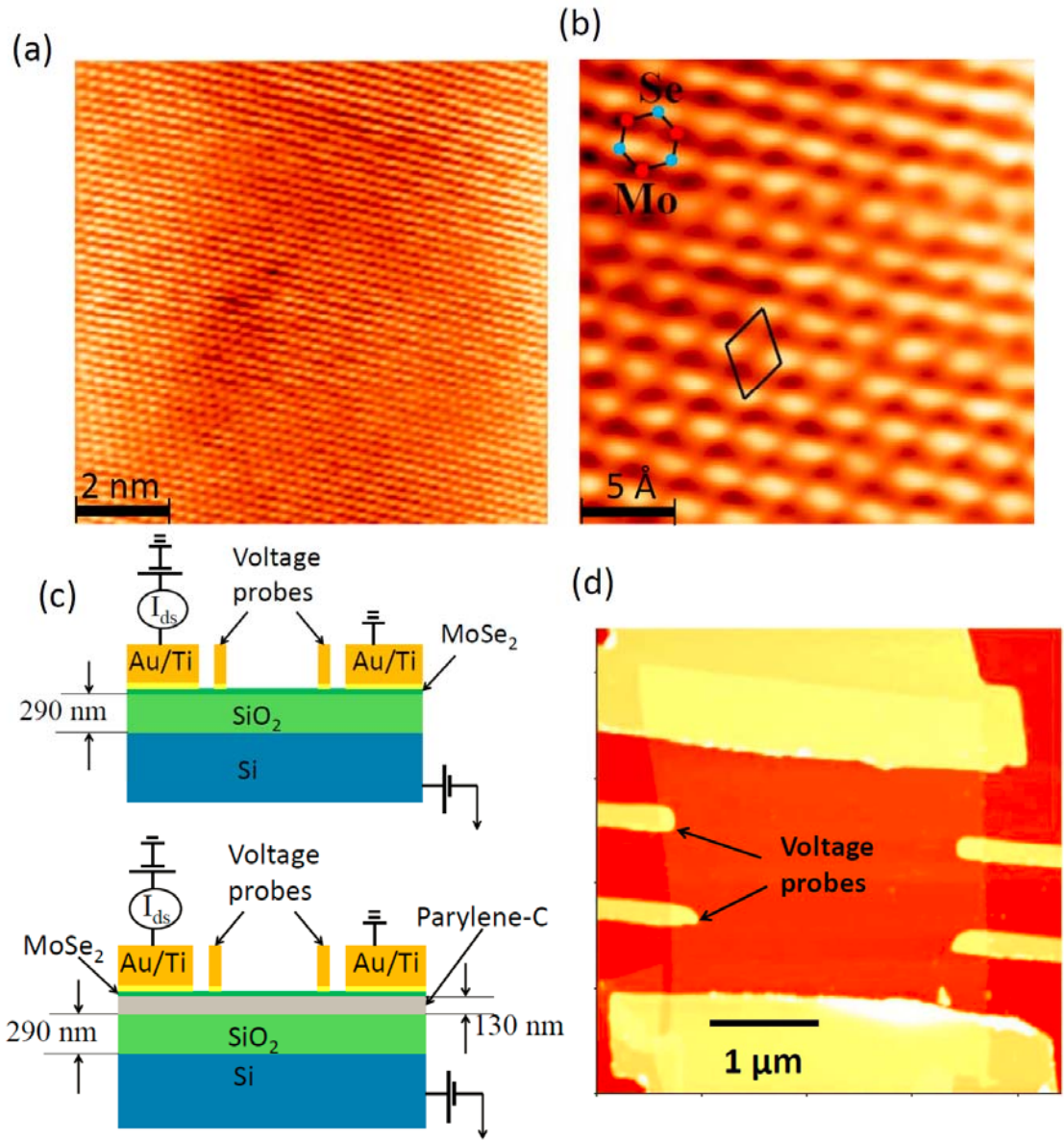

Figure 1.

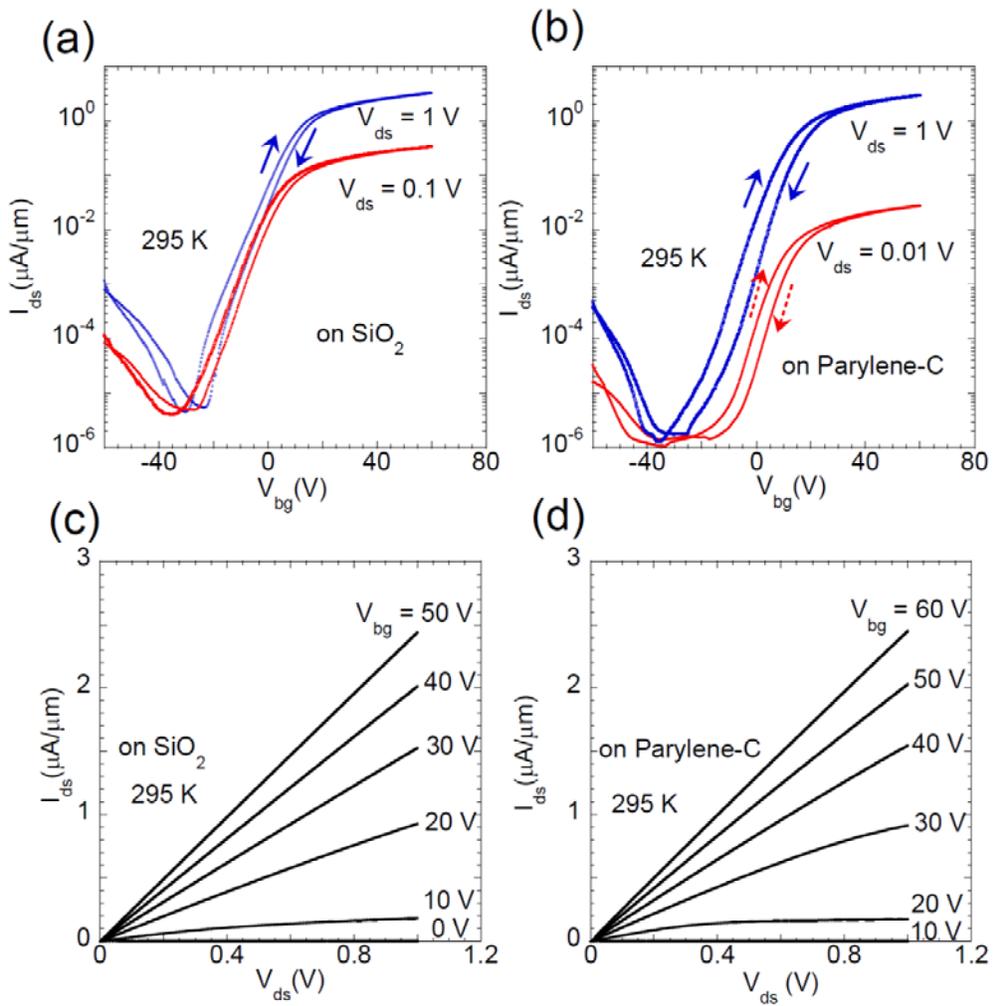

Figure 2.

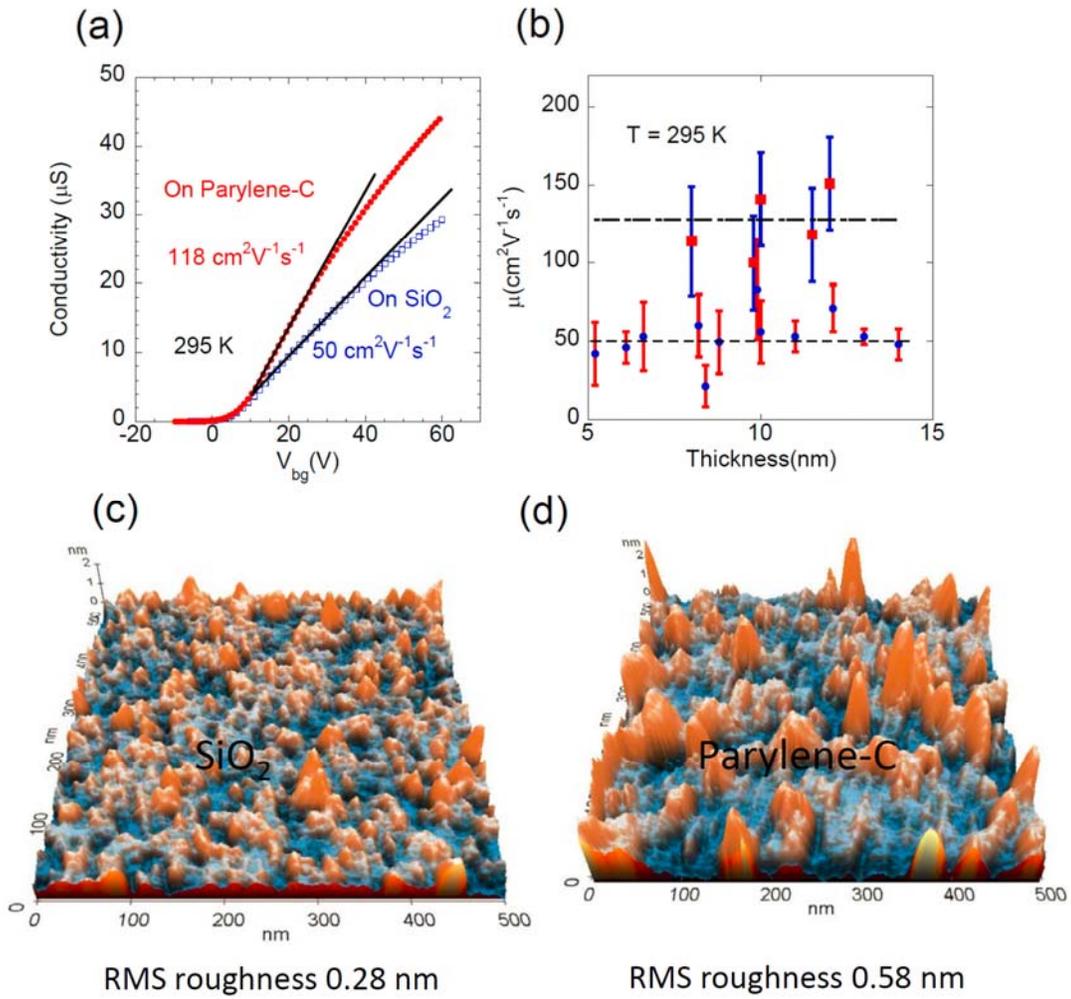

(a) On Parylene-C
118 cm²V⁻¹s⁻¹
295 K
On SiO₂
50 cm²V⁻¹s⁻¹

(b) T = 295 K

(c) SiO₂
RMS roughness 0.28 nm

(d) Parylene-C
RMS roughness 0.58 nm

Figure 3

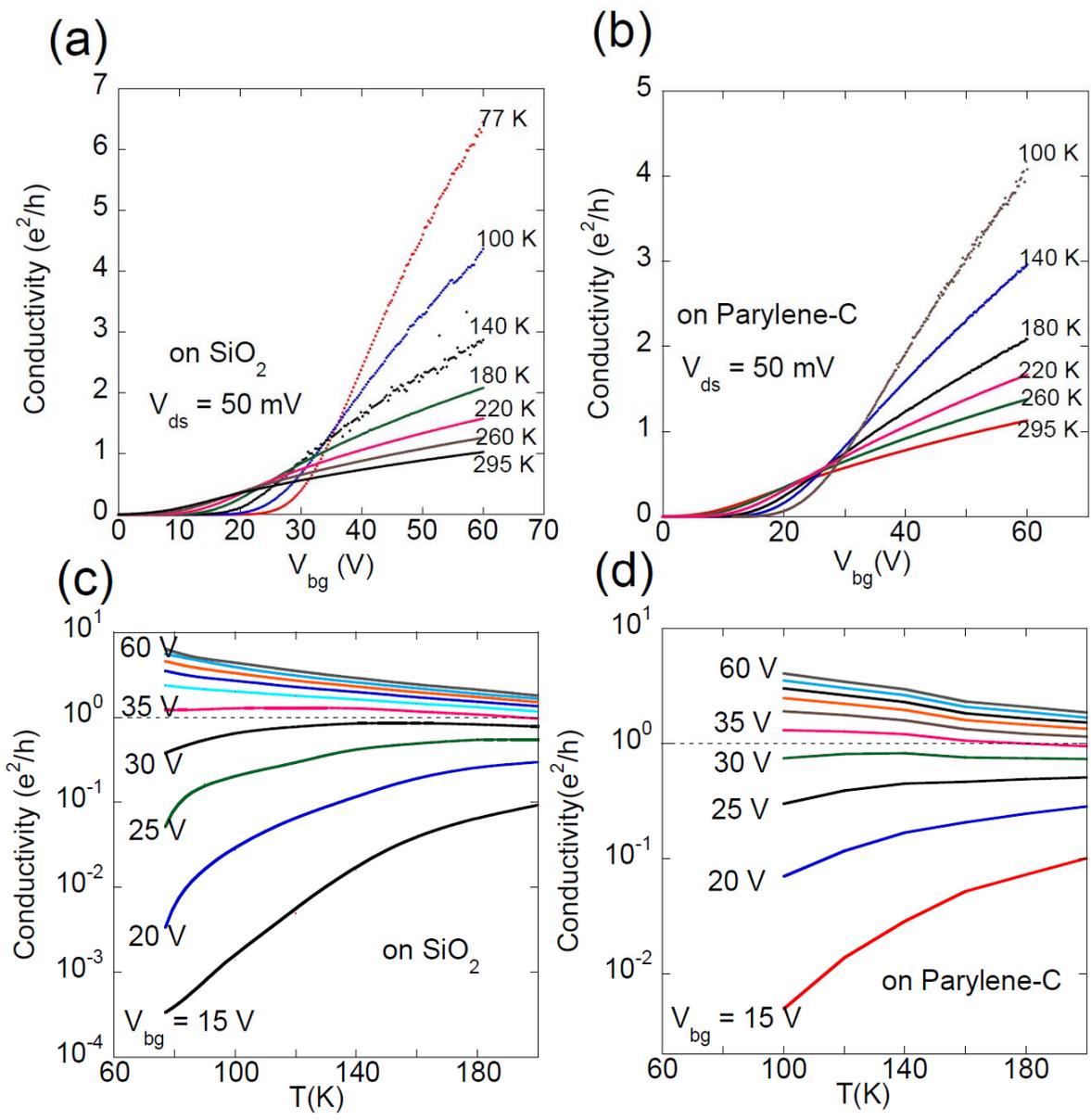

Figure 4

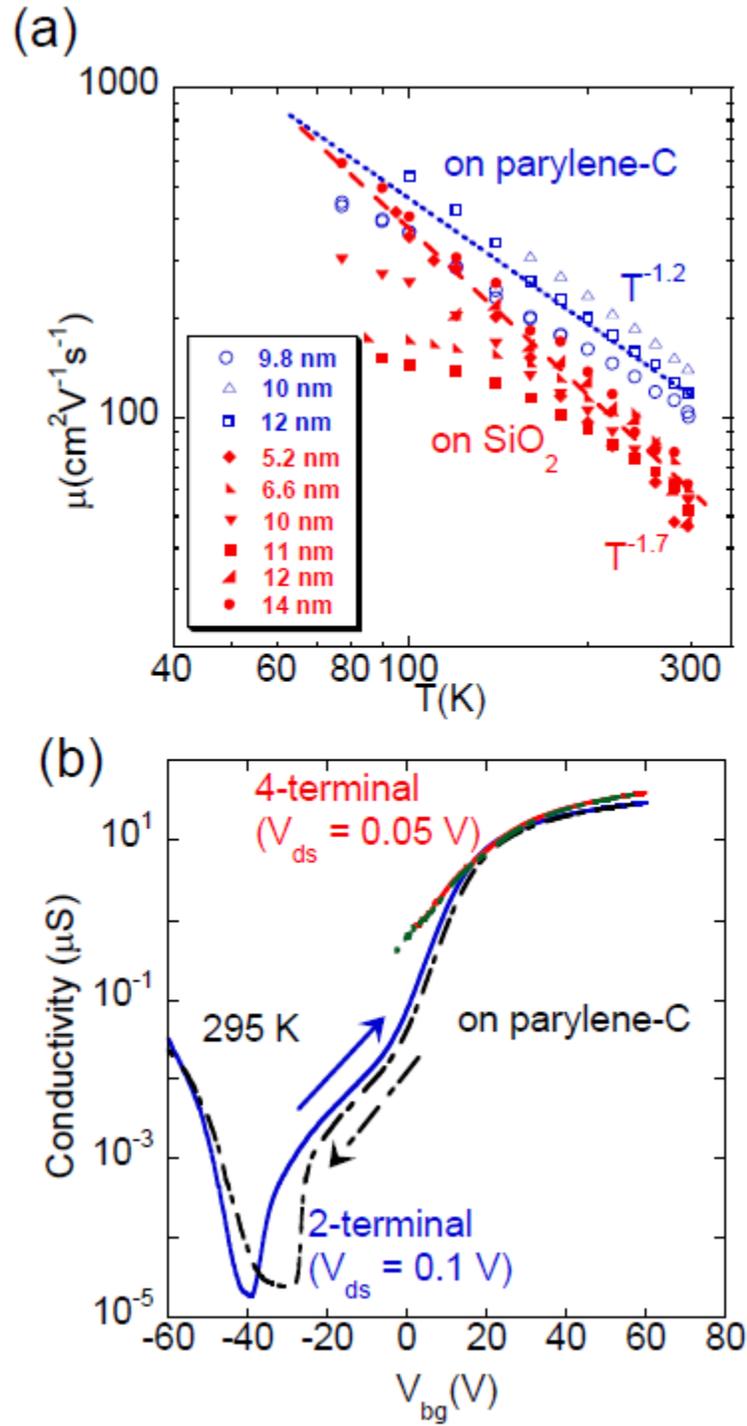

Figure 5

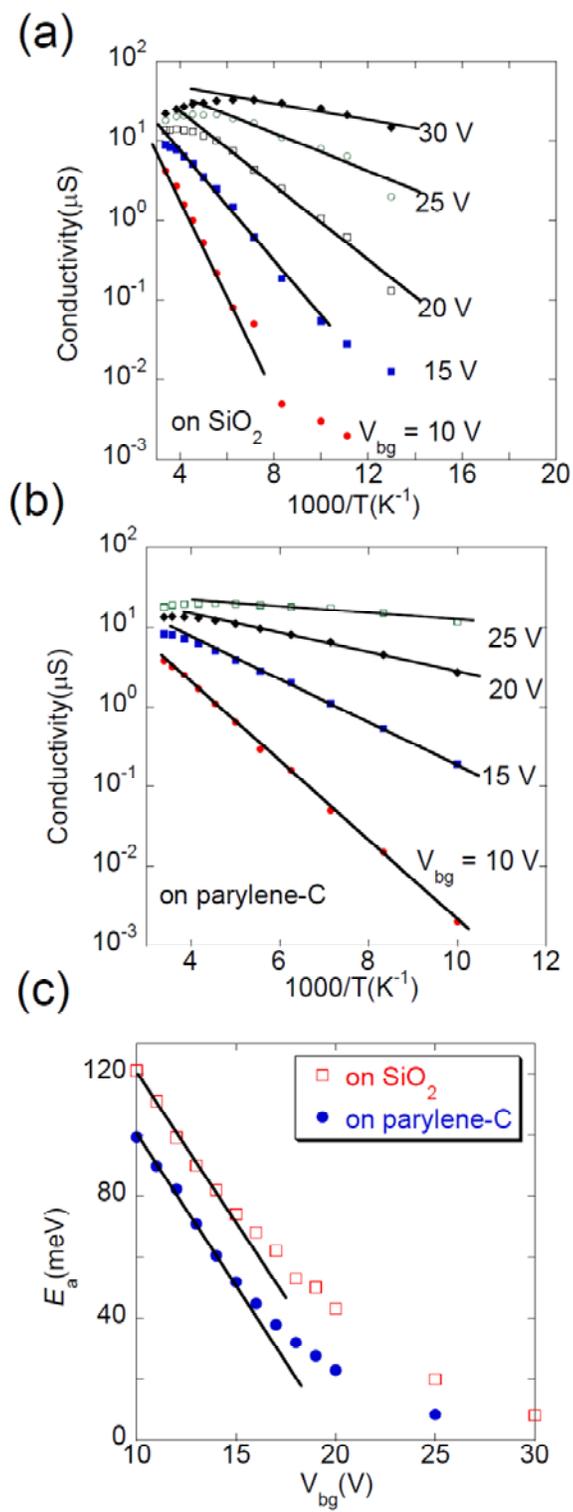

Figure 6

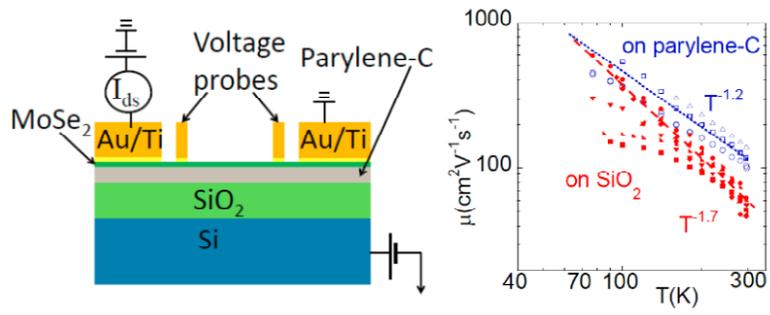

ToC graphic